\documentclass[preprint,12pt]{elsarticle}

\usepackage{amssymb}

\journal{Physica E}

\begin{document}

\begin{frontmatter}


\title{Hopping magnetoresistance in ion irradiated monolayer graphene}
\author[resnick]{I. Shlimak\corref{cor1}}
\ead{ishlimak@gmail.com}
\author[nano]{E. Zion}
\author[nano]{A.V. Butenko}
\author[resnick]{L. Wolfson}
\author[nano]{V. Richter}
\author[resnick]{Yu. Kaganovskii}
\author[nano]{A. Sharoni}
\author[engeneering]{A. Haran\fnref{fn1}}
\author[engeneering]{D. Naveh}
\author[resnick]{E. Kogan}
\author[resnick]{M. Kaveh}

\cortext[cor1]{Corresponding author}
\address[resnick]{Jack and Pearl Resnick Institute of Advanced Technology, Department of Physics, Bar-Ilan University, Ramat-Gan 52900, Israel}
\address[nano]{Institute of Nanotechnology and Advanced Materials, Bar-Ilan University, Ramat-Gan 52900, Israel}
\address[engeneering]{Faculty of Engineering, Bar-Ilan University, Ramat-Gan 52900, Israel}
\fntext[fn1]{permanent address: Soreq NRC, Yavne 8180000, Israel}





\begin{abstract}
Magnetoresistance (MR) of ion irradiated monolayer graphene samples with variable-range hopping (VRH) mechanism of conductivity was measured at temperatures down to $T = 1.8$ K in magnetic fields up to $B = 8$ T. It was observed that in perpendicular magnetic fields, hopping resistivity $R$ decreases, which corresponds to negative MR (NMR), while parallel magnetic field results in positive MR (PMR) at low temperatures. NMR is explained on the basis of the "orbital" model in which perpendicular magnetic field suppresses the destructive interference of many paths through the intermediate sites in the total probability of the long-distance tunneling in the VRH regime. At low fields, a quadratic dependence ($|\Delta R/R|\sim B^2$) of NMR is observed, while at $B > B^*$, the quadratic dependence is replaced by the linear one. It was found that all NMR curves for different samples and different temperatures could be merged into common dependence when plotted as a function of $B/B^*$. It is shown that $B^*\sim T^{1/2}$ in agreement with predictions of the "orbital" model. The obtained values of $B^*$ allowed also to estimate the localization radius $\xi$ of charge carriers for samples with different degree of disorder. PMR in parallel magnetic fields is explained by suppression of hopping transitions via double occupied states due to alignment of electron spins.
\end{abstract}



\end{frontmatter}


\section{Introduction}
\label{intro}

Defects are very useful tool to modify and control the physical properties of true two-dimensional (2d) material - monolayer graphene, therefore, disordered graphene attracts a lot of attention \cite{1,2,3,4,5,6,7}. The evolution of optical (Raman spectra) and electronic (conductivity) properties of graphene with increasing disorder has been investigated in many papers (see, for example, \cite{8,9,10,11}). It has been shown in [11] that increase of disorder leads to gradual change of the mechanism of conductivity from metallic one in pristine films to the regime of weak localization, and then to the mechanism of variable-range hopping (VRH) of strongly localized charge carriers.
In this work, we report the results of magnetoresistance (MR) measurements in samples of monolayer graphene strongly disordered with ion irradiation. MR in the VRH regime of conductivity has been observed earlier in different graphene-based structures: fluorinated graphene \cite{4}, graphene exposed to ozone \cite{5}, graphene oxide \cite{6}. In Ref. \cite{7}, MR was measured in monolayer graphene flakes subjected to Ga$^+$ ion irradiation. In one highly disordered sample, the negative MR was observed which was attributed to the crystalline-boundary scattering. In Ref. \cite{9}, disorder in graphene was introduced by C$^+$ ion irradiation with energy 35 keV. It was shown that at high dose of irradiation, the conductivity is described by the VRH mechanism, but no MR measurements were conducted. We are not aware about study of hopping MR in series of monolayer graphene samples with gradually increased degree of disorder.

\section{Experimental results and discussion}
\label{exp}

The investigated samples belong to a series of samples disordered by different dose of ion irradiation. All samples were fabricated by means of electron-beam lithography (EBL) on the common large-scale ($5\times 5$ mm) monolayer graphene film and divided into 6 groups. Initial sample before EBL was marked as sample 0. The samples from the first group, marked as sample 1, were not irradiated, while 5 others groups were subjected to different doses (from $5\times 10^{13}$ up to $1\times 10^{15}$ cm$^{-2}$) of irradiation with carbon ions with energy 35 keV. In Ref. \cite{10}, concentration of structural defects $N_D$ was determined for each group of samples using measurements of the Raman scattering. For samples 1 - 4, the values of $N_D$ in units of $10^{12}$ cm$^{-2}$ were 1, 3, 6 and 12 correspondingly. In Ref. \cite{11}, it was shown that the temperature dependence of resistivity $R(T)$ in initial sample 0 has a metallic character, while in slightly irradiated sample 1, conductivity is characterized by the regime of "weak localization". Measurements of MR in this sample showed a remarkable agreement with theoretical model \cite{12}. In more irradiated samples 2, 3 and 4, dependence $R(T)$ is described  by the variable-range hopping (VRH) mechanism of conductivity typical for strongly localized carriers [13].
As known, the form of $R(T)$ in VRH depends on the structure of the density-of-localized states $g(\mu)$ in the vicinity of the Fermi level (FL) $\mu$: when $g(\mu) \neq 0$, $R(T)$ is described by the "Mott VRH" which in 2d has the form of "$T^{– 1/3}$ - law":
\begin{eqnarray}
R(T) = R_0 \exp(T_M/T)^{1/3},\;\;\;  	T_M = C_M [g(\mu)\xi^2]^{-1	}.	
\end{eqnarray}
Here $R_0$ is a prefactor, $C_M = 13.8$ is the numerical coefficient, $\xi$ is the radius of localization.
The Coulomb interaction between localized carriers leads to their redistribution in the vicinity of the FL. This results in a "soft" Coulomb gap around FL, which in 2d has a linear form $g(\epsilon) = |\epsilon - \mu|(\kappa/e^2)$ where $\kappa$ is the dielectric constant of the material. In this case, $g(\mu) = 0$ and VRH is described by the "Efros-Shklovskii (ES) VRH" or "$T^{– 1/2}$ - law":
\begin{eqnarray}
R(T) = R_0\exp(T_{ES}/T)^{1/2},\;\;\;  	T_{ES} = C_{ES}(e^2/\kappa\xi)	
\end{eqnarray}
Here $C_{ES} = 2.8$ is the numerical coefficient.
Measurements of MR in samples 2, 3 and 4 was performed at temperatures down to 1.8 K in magnetic fields up to $B = 8$T in perpendicular  and in-plane (parallel) $B_{\parallel}$ geometry. It was observed that $B_{\perp}$ leads to negative magnetoresistance (NMR) while $B_{\parallel}$ results in positive magnetoresistance (PMR) at low temperatures, Fig. 1. This anisotropy shows unambiguously that MR in perpendicular and parallel fields has different origin: NMR is determined by the orbital mechanisms, while PMR is determined by the spin polarization. In that order we will discuss the results of measurements.

\begin{figure}[h]
\vskip -2cm
\begin{center}
\includegraphics[width= .5\columnwidth]{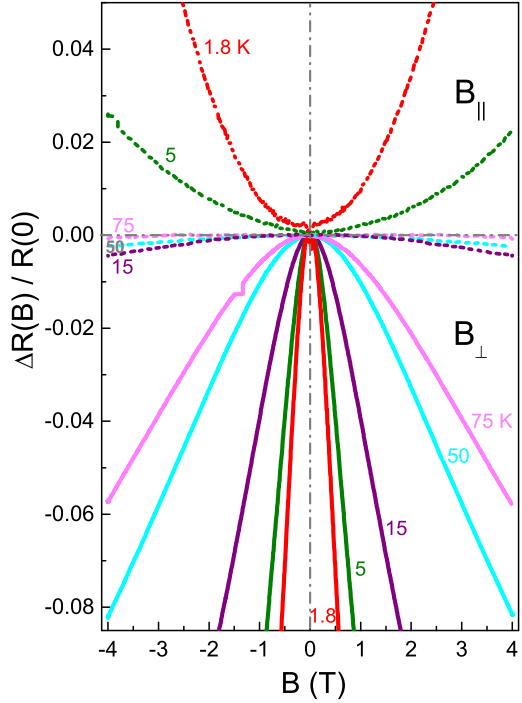}
\end{center}
\vskip -1cm
\caption{\label{fig:1}  $\Delta R/R$ as a function of $B$  for sample 3 in parallel and perpendicular magnetic fields at different temperatures in K, shown near each curve.}
\end{figure}

\section{NMR in perpendicular magnetic fields}
\label{mnr}

Fig. 2 shows the MR curves $\Delta R(B)/R(0)\equiv [R(B)-R(0)]/R(0)$ at different $T$ for all three samples on a linear scale. One can see, that NMR at fixed $T$ decreases with increase of disorder from sample 2 to 4. For samples 2 and 3, NMR increases with decreasing $T$. For sample 4, $\Delta R/R$ first increases with decreasing $T$, but at $T < 10$ K, NMR rapidly decreases and the curves seek to change the sign. It could be due to the standard positive MR caused by the shrinkage of the wave functions in perpendicular magnetic fields \cite{13}. Therefore we will discuss the NMR for sample 4 only down to 10 K.
In Fig. 3, NMR curves are plotted on the log-log scale.  On this scale, the slope to the curve is equal to the power $m$ in $\Delta R/R \sim B^m$. Quadratic dependence ($m = 2$) is observed at low fields up to some value $B^*$. At $B > B^*$, the quadratic dependence is replaced by linear one
($m = 1$) and then by sublinear dependencies. Some values of $B^*$ are shown in Fig. 3 by arrows.
Very weak effect of NMR (about 1-2\%) in VRH regime was earlier observed in three-dimensional (3d) conductivity in heavily doped and compensated Ge (for a review, see \cite{14}). In 2d, a significant NMR in the VRH regime has been observed in perpendicular magnetic fields in different systems [15-20]. Anisotropy of this effect in perpendicular and parallel fields unambiguously indicates the orbital nature of NMR, because effects due to spin polarization has to be isotropic. Theoretically the effect of orbital NMR in the VRH regime of conductivity has been discussed in \cite{21,22,23,24,25,26,27}. The main idea of the model suggested by Nguen, Spivak and Shklovskii \cite{21} is based on the following consideration.

\begin{figure}[h]
\begin{center}
\includegraphics[width= .5\columnwidth]{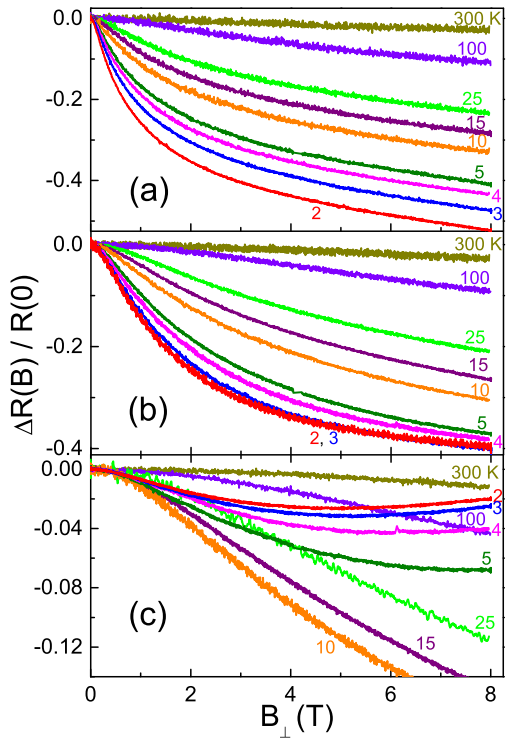}
\end{center}
\vskip -.7cm
\caption{\label{fig:2}  NMR of samples 2 (a), 3 (b) and 4(c) at different temperatures ($T$, K) shown near the curves. The density of structural defects $N_D$ (cm$^{-2}$): in sample 2 - $3\times 10^{12}$, 3  - $6\times 10^{12}$, 4 - $1.2\times 10^{13}$.}
\end{figure}

\begin{figure}[h]
\vskip -2cm
\begin{center}
\includegraphics[width= .5\columnwidth]{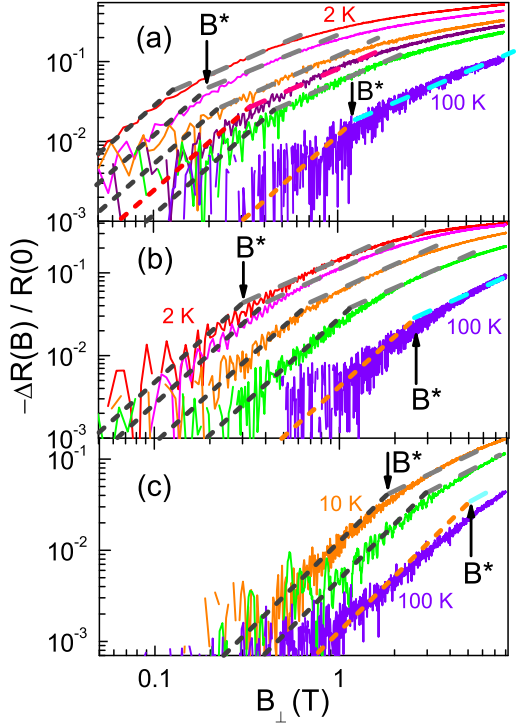}
\end{center}
\caption{\label{fig:3}  NMR as a function of B plotted on the log-log scale for samples 2 (a), 3 (b) and 4 (c). Short dashed lines correspond to
$m = 2$, dashed lines to $m = 1$. $B^*$ indicate the end of quadratic dependence.}
\end{figure}

In VRH, only part of localized states with energy levels within so-called "optimal band" around FL $\epsilon(T)$ are involved in the hopping process. In "Mott VRH", $\epsilon(T)$ decreases with decrease of temperature \cite{13}:
\begin{eqnarray}
\epsilon(T) = T^{2/3}[g(\mu) \xi^2]^{-1/3}				
\end{eqnarray}
Correspondingly the hopping distance $r_h$ increases, which gives $r_h \sim T^{-1/3}$:
\begin{eqnarray}
r_h \approx  [g(\mu) \epsilon(T)]^{-1/2} \approx \xi(T_M/T)^{1/3}.		
\end{eqnarray}
In "ES VRH", $r_h \sim T^{-1/2}$. Therefore, at low $T$, $r_h$ becomes much larger than the mean distance between localized centers, and the probability of the long-distance hop is determined by the interference of many paths of the tunneling through the intermediate sites which include scattering process (Fig. 4). All these scattered waves together with non-scattered direct wave contribute additively to the amplitude of the wave function $\Psi_{12}$ which reflects the probability for a charge carrier localized on site 1 to appear on site 2. There is no backscattering, and scattered waves decay exponentially with increasing distance as $\exp(–2r/\xi)$, therefore only the shortest paths contribute to $\psi_{12}$. All these paths are concentrated in a cigar-shaped domain of the length $r_h$, the width $D \approx (r_h \xi)^{1/2}$ and the area $A \approx r_h^{3/2}\xi^{1/2}$, where
$\alpha\leq  1$ is a numerical coefficient.

\begin{figure}[h]
\vskip -3cm
\begin{center}
\includegraphics[width= \columnwidth]{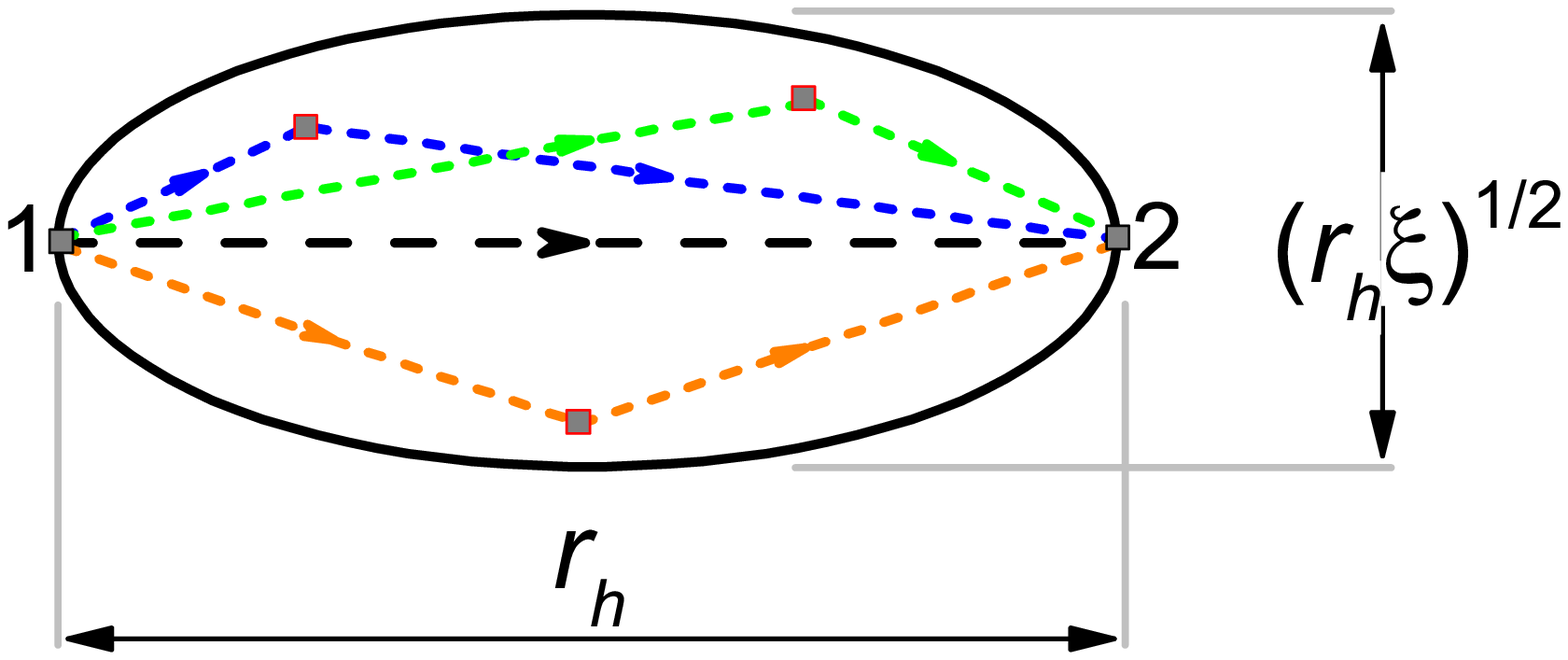}
\end{center}
\vskip -4cm
\caption{\label{fig:4}  Schematics of the cigar-shaped region with localized states contributing to the probability of an electron tunneling from center 1 to center 2.}
\end{figure}

As a result of averaging over different configurations, the contribution of the scattered sites to the total hopping probability vanishes due to destructive interference. The perpendicular magnetic field suppresses the interference which leads to the increase of the hopping probability and, therefore, to NMR. Taking into account that the hopping distance $r_h$ is fixed, the increase of the hopping probability could be considered as a small increase of the localization radius $\xi$. Both "Mott" and "ES" parameters $T_M$ and $T_{ES}$ depend on the value of $\xi$, but differently:
$T_M \sim \xi^{-2}$, Eq. (1), while $T_{ES} \sim \xi^{-1}$, Eq. (2). This gives rise to a conclusion that NMR has to be stronger for the case of "Mott VRH", then that for "ES VRH". This suggestion was observed in experiment. In Fig. 5(a), the temperature dependences of the modulus of NMR at fixed magnetic field $B = 4$ T for samples 2, 3 and 4 are shown. One can see that $\Delta R/R$ increases with decreasing $T$ linearly with $T^{-1/3}$ and  deviates from the straight line to smaller values at approximately the same temperatures where initial curves $R(T)$ at $B = 0$ deviate from $T^{–1/3}$ - law to the stronger $T^{–1/2}$ - law, Fig. 5(b).

\begin{figure}[h]
\vskip -2cm
\begin{center}
\includegraphics[width= .5\columnwidth]{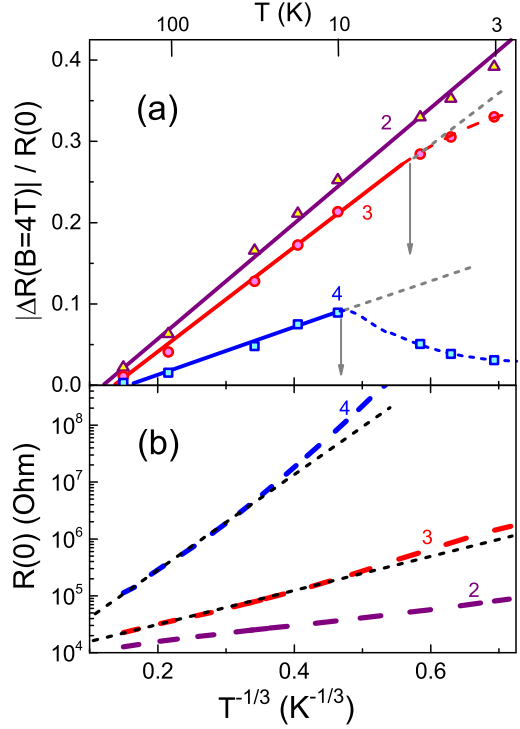}
\end{center}
\caption{\label{fig:5}  (a) – modulus of NMR at fixed magnetic field $B = 4$ T for samples 2, 3, 4 plotted as a function of $T^{-1/3}$. (b) - temperature dependence $R(T)$ for these samples at zero magnetic field.}
\end{figure}

In accordance with theoretical considerations \cite{21,22,23,24,25,26}, NMR as a function B has to be linear at moderate fields and quadratic at very low fields. In the "orbital" model, it is natural to normalize the magnetic field using the ratio $\eta = \Phi_B/\Phi_0$ where $\Phi_0 = h/2e \approx 2.07\times 10^{-15}$ W is the magnetic flux quantum and $\Phi_B = B\cdot A$ is the magnetic flux through the cigar-shape area. We suggest that $B^*$ corresponds to $\eta = 1$, i.e.
$B^* = \Phi_0/A \sim r_h^{-3/2}\zeta^{-1/2}$. Taking into account Eq. (4), one get ($\alpha\approx 1$):
\begin{eqnarray}
B^*= \Phi_0\xi^{-2}(T_M/T)^{-1/2}=\lambda T^{1/2},\;\;\;\;	\lambda = \Phi_0\xi^{-2}T_M^{-1/2}.		
\end{eqnarray}
In Fig. 6, the values of $B^*$ for all samples are plotted as a function of $T^{1/2}$. One can see that, indeed, $B^*\sim T^{1/2}$. Coefficient $\lambda$ is equal to 0.1, 0.24 and 0.58 [T K$^{-1/2}$] for samples 2, 3 and 4. Knowledge of $\lambda$ allows us to estimate with accuracy of $\alpha$ the values of localization radius $\xi$. For samples 2, 3 and 4, with $T_M=$ 68, 308 and 5960 K \cite{11}, this gives $\xi= 50, 22$ and $7$ nm correspondingly, which is quite reasonable: $\xi$ decreases with increase of disorder (from sample 2 to sample 4).

\begin{figure}[h]
\vskip -2cm
\begin{center}
\includegraphics[width= .7\columnwidth]{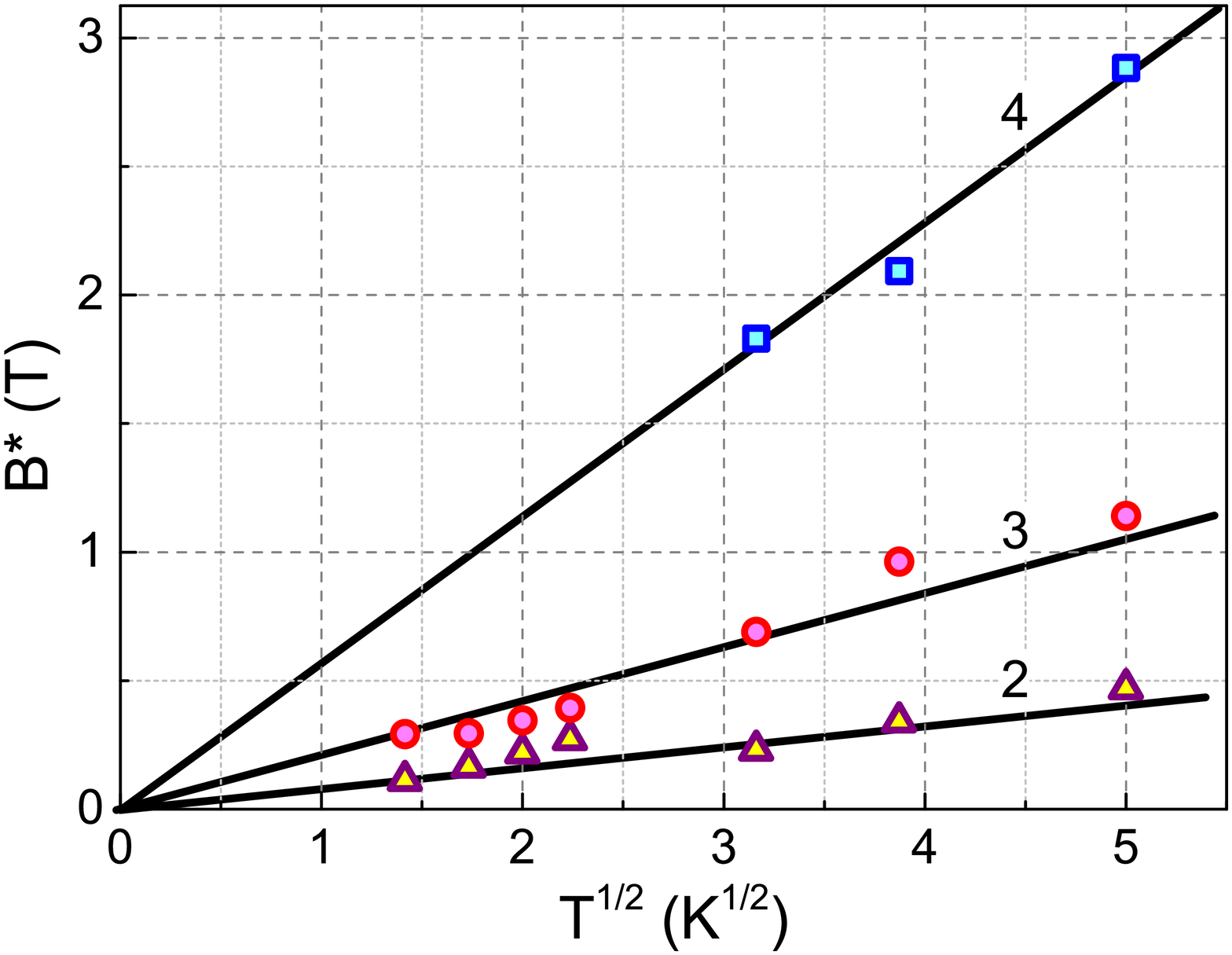}
\end{center}
\caption{\label{fig:6} The values of $B^*$ as a function of $T^{1/2}$. The sample numbers are shown near the straight lines. }
\end{figure}

We also use the values of $B^*$ in an attempt to merge the NMR data for all samples and all temperatures below 25 K. In Fig. 7, NMR curves from Fig. 2 are plotted as a function of dimensionless parameter $B/B^*$. One can see that all curves are merged in a universal dependence. No saturation of NMR is observed up to $\eta > 60$.

\begin{figure}[h]
\vskip -3cm
\begin{center}
\includegraphics[width= .8\columnwidth]{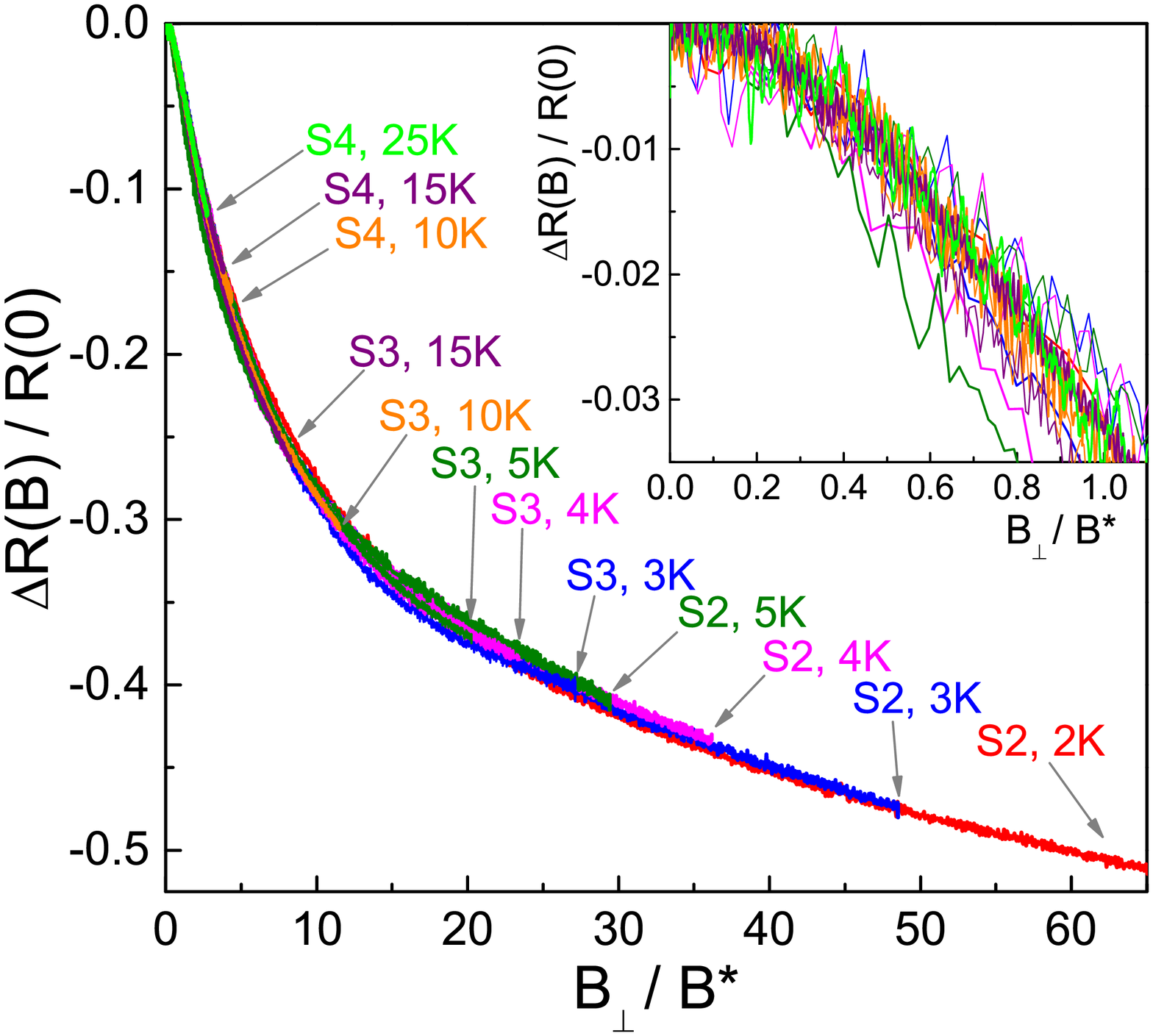}
\end{center}
\caption{\label{fig:7} The NMR data for different samples and different temperatures plotted as a function of dimensionless magnetic field  $B_{\perp}/B^*$. The arrows and numbers show the end of each curve and indicate the sample (2, 3, 4) and $T$. In insert, the NMR data are shown for small values  $B_{\perp}/B^*<1$. }
\end{figure}

\section{Positive MR in parallel magnetic fields}
\label{pmr}

In parallel magnetic fields, PMR is observed at low temperatures, Fig. 1. Very small NMR at high temperatures could be explained as the traces of NMR due to the possible folds on the surface of monolayer graphene film, where the parallel magnetic field has a perpendicular influence component. PMR appears only at low temperatures, at $T<5$ K and increases with decreasing $T$. PMR is proportional to $B_{\parallel}^2$ at low fields, and becomes linear with increasing $B$ (Fig. 8).

\begin{figure}[h]
\vskip -3cm
\begin{center}
\includegraphics[width= .8\columnwidth]{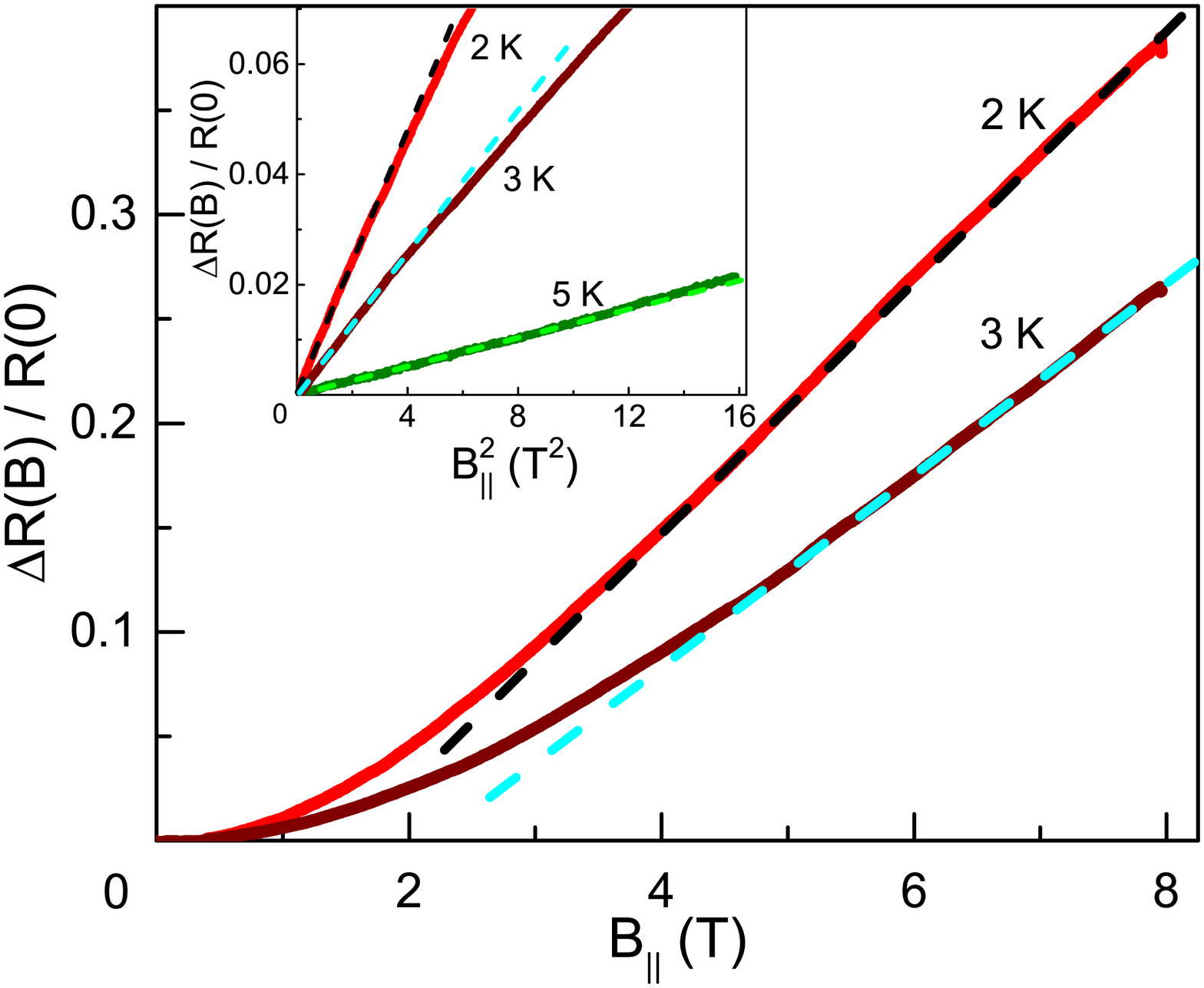}
\end{center}
\caption{\label{fig:8} PMR of sample 3 plotted as a function of $B_{\parallel}$. In insert, PMR data for low fields are plotted on a quadratic scale.}
\end{figure}

PMR in parallel magnetic fields has been observed earlier in 2d VRH regime in different systems: in Al$_{x}$In$_{1-x}$Sb/InSb quantum well \cite{15}, in a GaAs/Al$_x$Ga$_{1-x}$As  heterostructure \cite{27}.  Because the parallel magnetic field couples only to the electron spin, it means that the spin state of localized electrons influences the hopping conductivity despite the fact that it is not included explicitly in the expressions for VRH. The explanatory models of this effect are based on two ideas. The authors [28] considered the case when intermediate scattering centers (see Fig. 4) should be occupied to produce a negative scattering amplitude. Thus, the interference is depended on the mutual spin orientation of the hopping electron and electron localized on scattering center. In magnetic fields, all localized spins are aligned which increases the destructive interference and results in increase of the resistance. Another mechanism was suggested by Kurobe and Kamimura \cite{29} and studied in \cite{30}. In this model, it is recognized that a certain fraction of the states can accommodate two electrons. Double occupancy is possible if the on-site Coulomb repulsion $U$ between the electrons is smaller than the width of the energy distribution function of the localized states. It was already mentioned that in VRH,  only localized states with energy level within the narrow optimal band of width $\epsilon(T)$ around $\mu$ are involved in the hopping process, Eq. (3). However, for some states, which cannot participate in VRH at given temperature because the energy of the first electron $\epsilon^{(1)}$ is well below $\mu$, the energy of the second electron $\epsilon^{(2)} = \epsilon^{(1)} + U$ may be located just within the optimal band, Fig. 9. This allows those states to participate in the VRH at zero magnetic field. In strong field limit, all spins are polarized and, therefore, transitions through the double occupied states are suppressed which results in increase of resistance.

Fig. 1 shows that PMR is observed in our samples only at low temperatures. We think that this fact supports the mechanism [29] based on participation of the double occupied states, because the first mechanism \cite{28} has no limitation for observation at all temperatures. In the second mechanism, however, contribution of the double occupied states in VRH is important only when the width of the optimal band $\epsilon(T)$ which decreases with decreasing $T$, becomes less than $U$, otherwise in the case of an opposite inequality, $U \ll \epsilon(T)$, the localized states will either participate or not participate in VRH independently of the existence of the double occupied states. At moderate magnetic fields, theory \cite{30} predicts the linear dependence $\Delta R/R \sim (g_L\mu_BB)/T$, where $g_L$ is the Lande-factor and $\mu_B$ is the Bohr magneton, while at weak fields one expect the quadratic dependence $\Delta R/R\sim B^2$. This agrees with experiment (Fig. 8). Theory predicts also saturation PMR at strong fields when all electron spins are polarized. In our samples, no tendency to saturation was observed in magnetic fields up to 8 T.

 An interesting issue not discussed above is
a possible role of spin-orbit coupling for magnetoresistance in disordered graphene in the VRH regime.
The theory of anisotropic magnetoresistance of spin-orbit coupled carriers scattered
from polarized magnetic impurities, in which spin-orbit coupling enters via specific spin-textures on
the carrier Fermi surfaces and ferromagnetism via elastic scattering of carriers from polarized magnetic impurities, was proposed in Ref. \cite{31}.
Later, nonlinear anomalous Hall effect and negative magnetoresistance
in a system with random Rashba field was also studied theoretically \cite{32}. There it was shown that electron scattering from a fluctuating Rashba field in a two-dimensional
nonmagnetic electron system leads to a negative magnetoresistance arising solely due to spin-dependent
effects. 

However, knowing relative weakness of spin-orbit interaction in graphene,
we  think that it can only slightly modify the interpretation of the presented results.

\begin{figure}[h]
\vskip -1cm
\begin{center}
\includegraphics[width= .7\columnwidth]{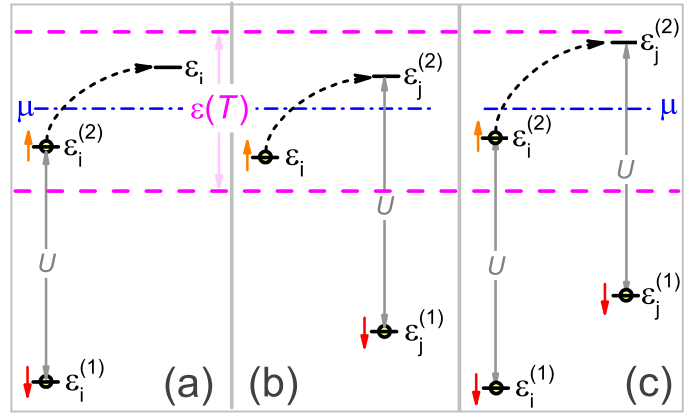}
\end{center}
\vskip -1cm
\caption{\label{fig:9} Schematic representation of the possible hopping transition via the double occupied states where the double occupied is an initial $\epsilon_i$ (a), final $\epsilon_j$ (b) or both initial and final states (c). $\mu$ represents the position of the Fermi level, dashed lines show the width $\epsilon(T)$ of the optimal band at given temperature, Eq. (3).}
\end{figure}

\section{Conclusions}

In conclusion, MR of strongly disordered monolayer graphene films with VRH mechanism of conductivity was measured in perpendicular $B_{\perp}$ and parallel $B_{\parallel}$ magnetic fields. It is shown that in perpendicular $B_{\perp}$, resistance decreases (negative MR), while $B_{\parallel}$ leads to positive MR at low temperatures. The NMR effect is explained in the framework of the "orbital" mechanism based on the interference of many paths through the intermediate sites in the probability of the long-distant tunneling in the VRH regime. Magnetic field $B^*$ at which the quadratic dependence $\Delta R/R \sim B^2$ is replaced by the linear one is determined as the merging parameter. It is shown that all MNR curves for different samples and different temperatures are merged in a universal dependence plotted as a function of dimensionless field $B/B^*$. It was shown that $B^*\sim T^{1/2}$ in accordance with the "orbital" model, the slope of this dependence allows us to estimate the localization radius $\xi$ for samples with different dose of irradiation. Expectedly, $\xi$ decreases with increase of disorder. The PMR effect in parallel fields could be explained by suppression in a strong magnetic field the hopping transitions via double occupied states due to electron spin polarization.

\end{document}